%% file: paper_NCE.tex
\documentclass[journal]{IEEEtran}

\ifCLASSINFOpdf
\else
   \usepackage[dvips]{graphicx}
\fi
\usepackage{url}
\usepackage[caption=false,font=scriptsize]{subfig}
\usepackage{graphicx}
\usepackage{siunitx}
\usepackage{tikz,adjustbox}

\usepackage{xspace}
\makeatletter
\DeclareRobustCommand\onedot{\futurelet\@let@token\@onedot}
\def\@onedot{\ifx\@let@token.\else.\null\fi\xspace}
\def\eg{\emph{e.g}\onedot}

\def\etal{\emph{et al}\onedot}
\makeatother

\usepackage{amsmath,amsfonts}
\usepackage{array}
\usepackage[caption=false,font=scriptsize]{subfig}
\usepackage{textcomp}
\usepackage{stfloats}
\usepackage{url}
\usepackage{verbatim}
\usepackage{graphicx}
\usepackage{cite}
\usepackage{stackengine}
\usepackage{xcolor,graphicx}
\usepackage{booktabs}

\usepackage{booktabs}
\usepackage{commath}
\usepackage{multirow}
\usepackage{amssymb}
\usepackage{siunitx}
\DeclareSIUnit\eps{eps}
\DeclareSIUnit\fps{fps}
\usepackage{amsthm}
\usepackage[ruled,linesnumbered,noend]{algorithm2e}
\usepackage{epstopdf}

\newcommand\exvis{120pt} 
\newcommand\intro{120pt}
\newcommand\exfeature{52pt}
\newcommand\exdetection{120pt}
\newcommand\exdenoising{120pt}

\usepackage{xspace}
\makeatletter
\DeclareRobustCommand\onedot{\futurelet\@let@token\@onedot}
\def\@onedot{\ifx\@let@token.\else.\null\fi\xspace}
\def\eg{\emph{e.g}\onedot}

\def\etal{\emph{et al}\onedot}
\makeatother
\usepackage[capitalize]{cleveref}
\crefname{section}{Sec.}{Secs.}
\Crefname{section}{Section}{Sections}
\Crefname{table}{Table}{Tables}
\crefname{table}{Tab.}{Tabs.}

\usepackage{fancyhdr}
\makeatletter
\let\ps@IEEEtitlepagestyle\ps@fancy
\makeatother
\begin{document}

\title{Event Encryption: Rethinking Privacy Exposure for Neuromorphic Imaging}
\author{Anonymous Submission}
\author{Pei Zhang, Shuo Zhu and Edmund Y. Lam,~\IEEEmembership{Fellow, IEEE}
\thanks{The authors are with the Department of Electrical and Electronic Engineering, The University of Hong Kong, Pokfulam, Hong Kong SAR, China (e-mail: zhangpei@eee.hku.hk, zhushuo@hku.hk, elam@eee.hku.hk). Edmund Y. Lam is also affiliated with ACCESS --- AI Chip Center for Emerging Smart Systems, Hong Kong Science Park, Hong Kong SAR, China.}%
\thanks{Corresponding author: Edmund Y. Lam.}}

\maketitle

\begin{abstract}
Bio-inspired neuromorphic cameras sense illumination changes on a per-pixel basis and generate spatiotemporal streaming events within microseconds in response, offering visual information with high temporal resolution over a high dynamic range. Such devices often serve in surveillance systems due to their applicability and robustness in environments with high dynamics and harsh lighting, where they can still supply clearer recordings than traditional imaging. In other words, when it comes to privacy-relevant cases, neuromorphic cameras also expose more sensitive data and pose serious security threats. Therefore, asynchronous event streams necessitate careful encryption before transmission and usage. This work discusses several potential attack scenarios and approaches event encryption from the perspective of neuromorphic noise removal, in which we inversely introduce well-crafted noise into raw events until they are obfuscated. Our evaluations show that the encrypted events can effectively protect information from attacks of low-level visual reconstruction and high-level neuromorphic reasoning, and thus feature dependable privacy-preserving competence. The proposed solution gives impetus to the security of event data and paves the way to a highly encrypted technique for privacy-protective neuromorphic imaging. 

\end{abstract}


\IEEEpeerreviewmaketitle

\section{Introduction}
\IEEEPARstart{N}{euromorphic} cameras encode per-pixel illumination changes of dynamic scenes with asynchronous events in microseconds, posing an imaging paradigm shift against conventional modalities that generate temporally-sparse, low-dynamic-range images~\cite{gallego2020event}. As such, these novel cameras, which feature high temporal precision and high dynamic range, can faithfully record visual information in environments with fast motion and harsh lighting.
\input{./src/intro}
Recently, neuromorphic cameras are being particularly regarded for the use of detection and recognition in surveillance fields. Thorny issues of privacy exposure inevitably arise when dealing with sensitive data. Earlier research believes that such an event-based solution can offer systematical guarantees for privacy and ethics considerations due to the incapability of capturing visible images~\cite{belbachir2012care}. However, advanced processing algorithms have been realizing accurate recognition, detection or reconstruction of gray-scale images from raw event streams. Besides, relative to the frame-based modality, neuromorphic imaging tends to expose more details since it can still offer blur-free recordings of dynamic objects under strong or weak illumination (as shown in~\Cref{fig:intro}). The resulting data exposure issues are thus receiving attention from the community. 

Neuromorphic privacy challenges were first raised in a recent study~\cite{du2021event}. In this work, we discuss more attack scenarios and approach event encryption from another perspective --- discrimination between events and noise. It is known that imaging slow-moving objects under low lighting can typically result in a great deal of noise that can obscure true visual information, and the noise is also hard to be eliminated~\cite{lichtsteiner2008128}. As such, encryption can be interpreted as a process of adding well-crafted noise into raw events for making human or machine reasoning failed. Accordingly, we propose a new solution for event encryption. It recursively synthesizes the noise with strong spatiotemporal correlation and high density estimation until the information is fully obfuscated. Such a method does not lead to the loss of the original data, enabling us to conceal sensitive biometric features in encryption and enjoy high-quality visualization after decryption. Our evaluations show that the processed events can defend against attacks from visualization, reconstruction, denoising and high-level reasoning, and thus feature reliable privacy-preserving ability. 

Inspired by frame-based image encryption, Du~\etal~\cite{du2021event} conducted the first investigation on this issue through incorporating two-dimensional chaotic mapping and a key updating mechanism. Compared with the counterpart, our technique enables direct encryption of four-dimensional event streams in an unsupervised manner, and also requires fewer operations for lossless decryption. Our research comprehensively studies the robustness of encrypted events to various neuromorphic analysis attacks. In addition to bringing faithful protection, the simple yet effective encryption is likely to be integrated with some existing noise filters.

\section{Related Work}
\textbf{Neuromorphic cameras} record local temporal contrast with microsecond temporal precision and over a $100$ \si{\decibel} dynamic range, and output asynchronous streams of events in response. Relative to frame-based cameras, they are more adaptive to highly dynamic scenes and less sensitive to illumination settings. Some neuromorphic cameras with an active pixel sensor embedded, such as DAVIS346~\cite{taverni2018front}, can also capture intensity frames at the same spatial resolution. In recent years, researchers have been trying to design event-driven solutions for various applications, such as auto-focusing~\cite{ge2023millisecond}, optical flow estimation~\cite{shiba22eccv} and motion detection~\cite{ge2022lens}, or to reconstruct high-quality images with the aid of event information~\cite{zhang2022unifying}.

The ongoing revolution of artificial intelligence technologies requires the support of an enormous quantity of visual data, whereas the raised \textbf{privacy} concerns have also been drawing public attention especially for the scenarios involving sensitive biometrics, such as facial expression tracking~\cite{becattini2022understanding} and gait recognition~\cite{wang2019ev}. Prior research is in favor of the privacy-preserving capability of neuromorphic cameras over traditional imaging. Nevertheless, existing techniques (\textbf{attacks}) can reconstruct proper representations of events that enable accurate machine or human reasoning. For example, purifying raw events for making them more informative~\cite{guo2022low}, or transforming streams into grid-based manifestations whereby one can interpret events in the form of images~\cite{gehrig2019end}, or performing inferences upon asynchronous events alone~\cite{perot2020learning}. These attacks put tremendous pressure on neuromorphic imaging from the security side.
 
A variety of \textbf{encryption} solutions are proposed for concealing sensitive visual information. Chaos-based variants leverage a pseudo-random sequence of numbers to shuffle spatial pixels~\cite{behnia2008novel}. Due to the fast parallel processing competence, optical techniques (\eg, nonlinear optics~\cite{hou2022image}, meta-hologram~\cite{qu2020reprogrammable}) are adopted in image cryptography. Compressive sensing~\cite{zhao2022deep} that is primarily used for data shrinking, also contributes to low-cost security enhancements~\cite{cambareri2015low}. Target to neuromorphic imaging, a recent proposal bridged traditional chaotic schemes with polarity flipping~\cite{du2021event}. This work continues the discussion on event encryption from a new perspective --- neuromorphic noise removal. 

\section{Methodology}
\subsection{Preliminary of Neuromorphic Imaging}
Neuromorphic cameras react to illumination changes with asynchronous streaming events~\cite{gallego2020event}, with each event being represented as
\begin{equation}
    \mathbf{e}_i = (\mathbf{x}_i, p_i, t_i),
\end{equation}
where $\mathbf{x}_i$ is a spatial pixel in which an event, indexed by $i$, is triggered at a certain timestamp $t_i$. The polarity $p_i \in \{-1, +1\}$ denotes the sign of the illumination change. Since a single event carries limited information, we gather a batch of events $\mathbf{E}$ over a time period for a more expressive representation of scenes
\begin{equation}
    \mathbf{E} = \{\mathbf{e}_i\}_{i=1:I} = \big\{(\mathbf{x}_i, p_i, t_i)\big\}_{i=1:I},
\end{equation}
where $I = |\mathbf{E}|$ is the number of events, and the timestamp increases monotonically with the index. 

Such an imaging modality is sensitive to various sources of interference, resulting in the output with a great amount of noise. Background Activity noise, which comes from junction leakage (in bright scenes) and thermal noise (in dark scenes), irregularly appears in the pixels without any activity~\cite{khodamoradi2018n,guo2022low}, whereas informative events, which are triggered by the edges of moving objects, typically have strong spatiotemporal correlation and polarity continuation~\cite{gallego2020event}. Exploiting the nearest-neighbor filter is a simple yet effective solution to discriminate between events and noise~\cite{czech2016evaluating}, which can be formulated as
\begin{equation}
    \mathbf{B}^{(i)} = \{j \in \mathbf{E} \mid f_{\mathbf{x}}(j, i) < T_{\mathbf{x}},~f_{t}(j, i) < T_{t},~f_{p}(j, i) = 0\},
\end{equation}
where $\mathbf{B}^{(i)}$ is a collection of the neighbors of an event $\mathbf{e}_i$. $f_{\mathbf{x}}$, $f_{t}$ and $f_{p}$ are the distance measures associated with the three dimensions of an event, and $T_{\mathbf{x}}$, $T_{t}$ are case-dependent thresholds. Such a filter only admits the events that sufficiently approach $\mathbf{e}_i$ in space-time while having the identical polarity. Besides, the number of events in a given space is normally far more than that of noise, leading to more densely clustered regions in which events reside~\cite{feng2020event}. 

Nevertheless, it still remains challenging for noise filters to have correct estimation when imaging slow-moving objects under low lighting, where the resulting events, similar to noise, have weak spatiotemporal correlation and sparse distribution, and the noise also accounts for a high proportion in quantity. In this particular case, visual information is significantly confused. As such, we rethink that \emph{whether event encryption can be regarded as an inverse process of event denoising, where a great amount of well-crafted noise is introduced into clean events until the information is totally obfuscated?} In terms of feasibility, well-established image encryption methods are not adapted to asynchronous streaming events due to their inability to handle temporal information. 
We call for a specific algorithm that already has sufficient theoretical support from neuromorphic noise removal. Similar to denoising, unsupervised event encryption without the assistance of auxiliary data can significantly increase the practicality. Moreover, as events encode incomplete visual information of scenes, the encryption and decryption should also be lossless to avoid the further signal loss. In what follows, we detail our methodology based on the above considerations.

\input{./src/encode}
\subsection{Event Encryption with Synthetic Noise}
By leveraging the features of events and noise, we propose an encryption algorithm (Algorithm~\ref{alg:encode}) by which we synthesize noise in the given space and disrupt the polarity continuation of the original events via a pseudo-random fashion.

The algorithm processes the raw event stream $\mathbf{E}$ to be the encrypted one $\tilde{\mathbf{E}}$. It first projects $\mathbf{E}$ into a two-dimensional plane $\mathbf{E}_\mathbf{x}$ that only has spatial information. The mask $\mathbf{M}$ represents the pixels in which noise exists, which is initialized by a function $\delta$ that can be either static (\eg, hard-coded rules) or adaptive (\eg, neural networks) as long as $\mathbf{M} \cap \mathbf{E}_\mathbf{x} = \varnothing$. For a position $\mathbf{x}_i$ where events exist, we compute its spatial neighbors $\mathbf{R} = \theta(\mathbf{x}_i, \mathbf{M})$ in the mask 
\begin{equation}\label{eq:r}
    \theta(\mathbf{x}_i, \mathbf{M}) = \{\mathbf{x} \in \mathbf{M} \mid f_{\mathbf{x}}(\mathbf{x}, \mathbf{x}_i) < T_{\mathbf{x}}\},
\end{equation}
by which the synthetic noise can spatially correlate with the events. The central events $\mathbf{E}_c$ contain a set of events triggered at the same position but at different timestamps. Then, given $\mathbf{R}$ and $\mathbf{E}_c$, we synthesize multiple noise events $\mathbf{N}$ by the function $\phi$ such that any one of them has the following attributes 
\begin{equation}\label{eq:n}
    \begin{cases} 
        \mathbf{x} \in \mathbf{R}\\ 
        p \in \{-1, +1\}\\
        t = t_i\left(1 + \sigma f_{\mathbf{x}}(\mathbf{x}, \mathbf{x}_i)\right)\\
        f_{t}(t, t_i) < T_{t}
    \end{cases},
\end{equation}    
where $p$ is randomized, $t$ relates to $t_i$ by the spatial distance between the noise and the event $\mathbf{e}_i$ under the constraint that they still temporally correlate with each other, and $\sigma$ is a scaling factor. The~\Cref{eq:r,eq:n} enforce that there is strong spatiotemporal correlation between the synthetic noise and original events. At each $\mathbf{x} \in \mathbf{R}$, we also allows $|\mathbf{N}| = |\mathbf{E}_c|$ whereby the noise shares the same density estimate with the events. The algorithm recursively fills the space with noise and terminates when all the pixels in the mask have been traversed. We elaborate more on the processing in~\Cref{fig:method}. Finally, a reversible mapping $\lambda$ disrupts the polarity continuation
\begin{equation}\label{eq:p}
    \lambda(\tilde{\mathbf{E}}) = \{\mathbf{e} \in \tilde{\mathbf{E}} \mid p = p \oplus (\texttt{szudzik}(\mathbf{x}) \bmod 2)\},
\end{equation}
where $\oplus$ is an XOR gate, and \texttt{szudzik} is a pairing function that uniquely encodes a two-dimensional value into an integer~\cite{szudzik2006elegant}. Since there is a setting $\mathbf{M} \cap \mathbf{E}_\mathbf{x} = \varnothing$ in the algorithm, the \texttt{key} to the decoding can simply be the encrypted $\mathbf{E}_\mathbf{x}$
\begin{equation}
    \texttt{key} = \rho \left(\texttt{szudzik}(\mathbf{E}_\mathbf{x})\right),
\end{equation}
where $\rho$ can be any well-established encryption method that can handle a list of numbers. Therefore, in the decoding phase $\tilde{\mathbf{E}} \mapsto \mathbf{E}$, we can use the \texttt{key} to locate the pixels where true events exist and~\Cref{eq:p} to retrieve the original polarity.
\input{./src/method}

\section{Experiments}
\subsection{Experimental Settings}
In our experiments on Algorithm~\ref{alg:encode}, the function $\delta$ computes the mask $\mathbf{M} = \mathbf{S} \setminus \mathbf{E}_\mathbf{x}$, where $\mathbf{S}$ is the camera resolution. The measures $f_{\mathbf{x}}$, $f_{t}$ and $f_{p}$ calculate the $L_1$ distance, while the spatial threshold $T_{\mathbf{x}}$ is set to $1$. The scaling factor $\sigma$ is configured as $0.05$, and the value of the temporal threshold $T_{t}$ thus varies with the attributes of the events to be encrypted. The evaluations that follow are conducted based on several public datasets~\cite{mueggler2017event,orchard2015converting,bi2019graph,perot2020learning,de2020large}. To alleviate the influence of noise inherent in the original data on task performance, we exclude the samples where noise predominates in quantity. Raw event streams are encrypted using the algorithm with the given settings and then serve as the input of downstream applications (attacks).

\subsection{Attacks from Visualization and Reconstruction}
\input{./src/ex_vis}

Neuromorphic imaging records scene brightness changes in a set of tuples and was once considered privacy-preserving. Nevertheless, events can be visualized in a two-dimensional plane by counting or accumulation. Moreover, they can also bring images into being high-dynamic-range states such that more sensitive details are exposed. \Cref{fig:ex_vis} shows an example based on DAVIS 240C Datasets~\cite{mueggler2017event}. The raw events, which are visualized in the form of an event frame~\cite{rebecq2017real}, can offer more facial information than the underexposed image, and the associated reconstruction~\cite{scheerlinck2018continuous} can even deliver a high-quality image that exposes sensitive biometrics. Our encrypted data conceal all the informative features and enable the reconstruction malfunctioned, supplying sufficient privacy protection.

\subsection{Attacks from Neuromorphic Noise Removal}
\input{./src/ex_denoising}
Neuromorphic imaging yields noise in response to interference, and the denoising is often used to purify raw events for making them more informative. Our encryption approach aims to fake the noise that denoising algorithms cannot suppress. In~\Cref{fig:ex_stats}, we study whether the encrypted events can resist the attacks from existing methods, such as Zhang~\etal~\cite{zhang2023neuro}, Wu~\etal~\cite{wu2020probabilistic}, Feng~\etal~\cite{feng2020event} and the nearest-neighbor filter (NNf)~\cite{czech2016evaluating}. For comparison, we allow the events filled with random noise to be another set of input, and keep the signal-to-noise ratio (SNR) of the two input data sets the same (SNR = $1$). The figure shows that the denoising methods can remove random noise and raise the SNR, but their effects on the encrypted events are fairly negligible, indicating that they cannot distinguish the true events from our fake noise.

\input{./src/ex_recognition}

\subsection{Attacks from High-level Neuromorphic Reasoning}
One can use advanced techniques to perform high-level reasoning on event streams and then efficiently acquire privacy-relevant information. Evaluated on N-MNIST~\cite{orchard2015converting} and ASL-DVS~\cite{bi2019graph}, \Cref{tab:ex_recognition} presents the top-1 recognition accuracy on raw events, on raw events with $50\%$ random noise, on the encrypted events by the competitor (Du~\etal~\cite{du2021event}) and on our encrypted events. The approaches, which are either grid-based~\cite{rebecq2017real,gehrig2019end,wang2022exploiting} or graph-based~\cite{li2021graph}, can still have good performance when the data are filled with random noise, but they fail for recognition on the encrypted ones. In~\Cref{fig:ex_feature}, we diagnose the learning process and the feature responses of a trained network. It shows that the network cannot learn any meaningful information from the encrypted events. 

With similar settings on tested samples, \Cref{tab:ex_detect} investigates neuromorphic object detection on 1 MEGAPIXEL~\cite{perot2020learning} and GEN1~\cite{de2020large}, and \Cref{fig:ex_detection} gives the visualization. The model can correctly detect \texttt{person} or \texttt{car} on the raw events, but it gives wrong results when the input is encrypted. The above experiments prove that our encryption can drastically alter the underlying pattern of events and make it more difficult for learning, and can even lead to more degraded performance compared with the counterpart.

\input{./src/ex_feature}
\input{./src/ex_detect}
\input{./src/ex_detection}

\section{Limitation and Discussion}
While our approach exhibits satisfactory performance in the evaluations described, it is essential to recognize the constraints. The algorithm is a procedure with at least quadratic time complexity, and non-deterministic runtime, which is affected by several factors including the sample to be encrypted, camera resolution as well as the noise mask used, can intolerably increase as the mask covers a large area of pixels. In addition, the proposed encryption fails to be lossless in the absence of a decryption key, posing a need for further enhancements.

\section{Conclusion}
Neuromorphic imaging can offer clear recordings of dynamic objects and is thus particularly practical in surveillance systems. However, similar to frame-based modalities, it also suffers from thorny privacy exposure issues. To enhance the security level of event data, we develop an encryption method that can synthesize pseudo-random noise such that all the visual information is obfuscated. Evaluations show that our encrypted events can resist the attacks from visualization, reconstruction, denoising and high-level reasoning, and thus have robust privacy-preserving ability. Nevertheless, in terms of execution time and energy consumption, the overhead of computing on encrypted data has not been fully investigated in this work. This emerging field is still requiring more attention and further research.

\section{Acknowledgments}
This work was supported in part by the Research Grants Council of Hong Kong SAR (GRF 17201620, 17200321) and by ACCESS --- AI Chip Center for Emerging Smart Systems, sponsored by InnoHK funding, Hong Kong SAR. The authors have confirmed that any identifiable participants in this study have given their consent for publication.

\bibliographystyle{IEEEtran}
\bibliography{refs}

\end{document}

%% file: src/intro.tex
\begin{figure}[t]
    \centering
    \subfloat[]{\frame{\includegraphics[width=\intro]{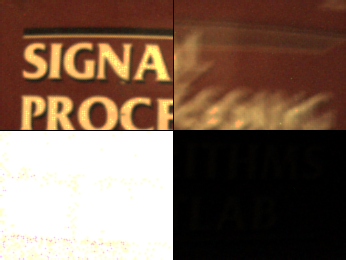}}}\hspace{2pt}
    \subfloat[]{\frame{\includegraphics[width=\intro]{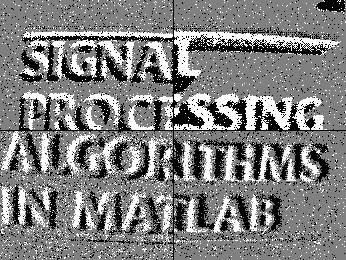}}}
    \caption{(a) When the intensity image (upper left) suffers from heavy motion blur (upper right) and being overexposed (lower left) or underexposed (lower right), it loses significant details and cannot deliver informative visualization. (b) Neuromorphic cameras are robust to harsh illumination conditions, such that the corresponding events still have clear recordings with high-temporal-precision and high-dynamic-range states, where sensitive information is also prone to be exposed.}
    \label{fig:intro}
\end{figure}

%% file: src/encode.tex
\begin{algorithm}[t]
    \caption{Event Encryption}
    \label{alg:encode}
    \SetKwInOut{Input}{Input}
    \SetKwInOut{Output}{Output}
    \SetKw{Continue}{continue}
    \SetKw{Break}{break}
    \DontPrintSemicolon
    \Input{$\mathbf{E}$}
    \Output{$\tilde{\mathbf{E}}$}
    Initialize $\mathbf{E}_\mathbf{x} = \{\mathbf{x}_i\}_{i=1:I}$ from $\mathbf{E}$, and  $\tilde{\mathbf{E}} = \mathbf{E}$;\;
    Initialize \texttt{mask} $\mathbf{M}$ via \texttt{function} $\delta$\;
    \ForEach{$\mathbf{x}_i \in \mathbf{E}_\mathbf{x}$}
    {
        Find \texttt{spatial neighbors} $\mathbf{R} = \theta(\mathbf{x}_i, \mathbf{M})$\;
        \lIf{$\mathbf{R} = \varnothing$}
        {
            \Continue
        }
        Find \texttt{central events} $\mathbf{E}_c = \{\mathbf{e} \in \mathbf{E} \mid \mathbf{x} = \mathbf{x}_i\}$\;
        Synthesize \texttt{noise} $\mathbf{N} = \phi(\mathbf{R}, \mathbf{E}_c)$\;
        $\tilde{\mathbf{E}} = \tilde{\mathbf{E}} \cup \mathbf{N}$\;
        $\mathbf{E}_\mathbf{x} = \mathbf{E}_\mathbf{x} \cup \mathbf{R}$\;
        $\mathbf{M} = \mathbf{M} \setminus \mathbf{R}$\;
        \lIf{$\mathbf{M} = \varnothing$}
        {
            \Break
        }
    }
    Compute \texttt{polarity mapping} $\tilde{\mathbf{E}} = \lambda(\tilde{\mathbf{E}})$
\end{algorithm}

%% file: src/method.tex
\begin{figure}[t]

\begin{center}

\tikzset{every picture/.style={line width=0.75pt}} %set default line width to 0.75pt        

\begin{tikzpicture}[x=0.75pt,y=0.75pt,yscale=-1,xscale=1]

%Shape: Grid [id:dp4674659138729642] 
\draw  [draw opacity=0] (211,110) -- (301,110) -- (301,200) -- (211,200) -- cycle ; \draw   (241,110) -- (241,200)(271,110) -- (271,200) ; \draw   (211,140) -- (301,140)(211,170) -- (301,170) ; \draw   (211,110) -- (301,110) -- (301,200) -- (211,200) -- cycle ;
%Shape: Square [id:dp5311018443105515] 
\draw  [fill={rgb, 255:red, 80; green, 227; blue, 194 }  ,fill opacity=0.55 ] (241,140) -- (271,140) -- (271,170) -- (241,170) -- cycle ;
%Shape: Square [id:dp5666708615995888] 
\draw  [fill={rgb, 255:red, 74; green, 144; blue, 226 }  ,fill opacity=0.55 ] (271,140) -- (301,140) -- (301,170) -- (271,170) -- cycle ;
%Shape: Square [id:dp34324661383284105] 
\draw  [fill={rgb, 255:red, 74; green, 144; blue, 226 }  ,fill opacity=0.55 ] (211,140) -- (241,140) -- (241,170) -- (211,170) -- cycle ;
%Shape: Square [id:dp85588877054911] 
\draw  [fill={rgb, 255:red, 74; green, 144; blue, 226 }  ,fill opacity=0.55 ] (241,110) -- (271,110) -- (271,140) -- (241,140) -- cycle ;
%Shape: Square [id:dp11955175475479907] 
\draw  [fill={rgb, 255:red, 74; green, 144; blue, 226 }  ,fill opacity=0.55 ] (241,170) -- (271,170) -- (271,200) -- (241,200) -- cycle ;
%Shape: Grid [id:dp6600367724043636] 
\draw  [draw opacity=0] (351,110) -- (441,110) -- (441,200) -- (351,200) -- cycle ; \draw   (381,110) -- (381,200)(411,110) -- (411,200) ; \draw   (351,140) -- (441,140)(351,170) -- (441,170) ; \draw   (351,110) -- (441,110) -- (441,200) -- (351,200) -- cycle ;
%Shape: Square [id:dp6854261385620974] 
\draw  [fill={rgb, 255:red, 80; green, 227; blue, 194 }  ,fill opacity=0.55 ] (381,140) -- (411,140) -- (411,170) -- (381,170) -- cycle ;
%Shape: Square [id:dp780554669078218] 
\draw  [fill={rgb, 255:red, 74; green, 227; blue, 194 }  ,fill opacity=0.55 ] (411,140) -- (441,140) -- (441,170) -- (411,170) -- cycle ;
%Shape: Square [id:dp03022560886301151] 
\draw  [fill={rgb, 255:red, 74; green, 227; blue, 194 }  ,fill opacity=0.55 ] (351,140) -- (381,140) -- (381,170) -- (351,170) -- cycle ;
%Shape: Square [id:dp5178986785413988] 
\draw  [fill={rgb, 255:red, 74; green, 227; blue, 194 }  ,fill opacity=0.55 ] (381,110) -- (411,110) -- (411,140) -- (381,140) -- cycle ;
%Shape: Square [id:dp23413552942359284] 
\draw  [fill={rgb, 255:red, 74; green, 227; blue, 194 }  ,fill opacity=0.55 ] (381,170) -- (411,170) -- (411,200) -- (381,200) -- cycle ;
%Shape: Square [id:dp8738078730511669] 
\draw  [fill={rgb, 255:red, 80; green, 144; blue, 226 }  ,fill opacity=0.55 ] (351,110) -- (381,110) -- (381,140) -- (351,140) -- cycle ;
%Shape: Square [id:dp024326548250706015] 
\draw  [fill={rgb, 255:red, 80; green, 144; blue, 226 }  ,fill opacity=0.55 ] (351,170) -- (381,170) -- (381,200) -- (351,200) -- cycle ;
%Shape: Square [id:dp8248900866843556] 
\draw  [fill={rgb, 255:red, 80; green, 144; blue, 226 }  ,fill opacity=0.55 ] (411,170) -- (441,170) -- (441,200) -- (411,200) -- cycle ;
%Shape: Square [id:dp05796854604419066] 
\draw  [fill={rgb, 255:red, 80; green, 144; blue, 226 }  ,fill opacity=0.55 ] (411,110) -- (441,110) -- (441,140) -- (411,140) -- cycle ;

% Text Node
\draw (389,151) node [anchor=north west][inner sep=0.75pt]   [align=left] {$\mathbf{E}_c$};
% Text Node
\draw (383,116) node [anchor=north west][inner sep=0.75pt]   [align=left] {$\mathbf{N}^{(1)}_1$};
% Text Node
\draw (413,146) node [anchor=north west][inner sep=0.75pt]   [align=left] {$\mathbf{N}^{(1)}_2$};
% Text Node
\draw (353,146) node [anchor=north west][inner sep=0.75pt]   [align=left] {$\mathbf{N}^{(1)}_4$};
% Text Node
\draw (249,151) node [anchor=north west][inner sep=0.75pt]   [align=left] {$\mathbf{E}_c$};
% Text Node
\draw (243,116) node [anchor=north west][inner sep=0.75pt]   [align=left] {$\mathbf{N}^{(1)}_1$};
% Text Node
\draw (273,146) node [anchor=north west][inner sep=0.75pt]   [align=left] {$\mathbf{N}^{(1)}_2$};
% Text Node
\draw (243,176) node [anchor=north west][inner sep=0.75pt]   [align=left] {$\mathbf{N}^{(1)}_3$};
% Text Node
\draw (213,146) node [anchor=north west][inner sep=0.75pt]   [align=left] {$\mathbf{N}^{(1)}_4$};
% Text Node
\draw (383,176) node [anchor=north west][inner sep=0.75pt]   [align=left] {$\mathbf{N}^{(1)}_3$};
% Text Node
\draw (413,176) node [anchor=north west][inner sep=0.75pt]   [align=left] {$\mathbf{N}^{(2)}_3$};
% Text Node
\draw (413,116) node [anchor=north west][inner sep=0.75pt]   [align=left] {$\mathbf{N}^{(2)}_2$};
% Text Node
\draw (353,176) node [anchor=north west][inner sep=0.75pt]   [align=left] {$\mathbf{N}^{(2)}_4$};
% Text Node
\draw (353,116) node [anchor=north west][inner sep=0.75pt]   [align=left] {$\mathbf{N}^{(2)}_1$};

\end{tikzpicture}
\end{center}
\vspace{-4pt}
\caption{Illustrations of our recursive encryption algorithm. There is a two-dimensional plane with $3 \times 3$ pixels, where multiple events $\mathbf{E}_c$ are triggered in the center, and the rest of the pixels are on the mask for synthetic noise. In the $1$st layer of the recursion (left), the algorithm synthesizes the noise $\mathbf{N}^{(1)}_i (i=1,2,3,4)$ in $4$ spatial neighbors horizontally/vertically adjacent to $\mathbf{E}_c$, where $|\mathbf{N}^{(1)}_i| = |\mathbf{E}_c|$. The resulting $\mathbf{N}^{(1)}_i \cup \mathbf{E}_c$ will be the input of the $2$nd layer of the recursion (right). The algorithm, which is blind to $\mathbf{N}^{(1)}_i$ and $\mathbf{E}_c$, synthesizes the noise $\mathbf{N}^{(2)}_i (i=1,2,3,4)$ based on the adjacent events.}
\label{fig:method}
\end{figure}
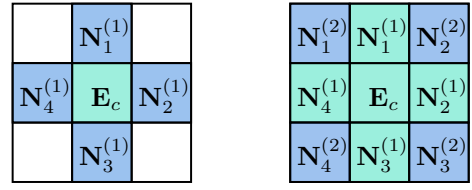

%% file: src/ex_vis.tex
\begin{figure*}[t]
    \centering
    \subfloat[Raw Events]{\frame{\includegraphics[width=\exvis]{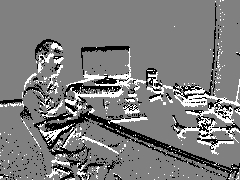}}}
    \hspace{6pt}%
    \subfloat[Underexposed Image]{\frame{\includegraphics[width=\exvis]{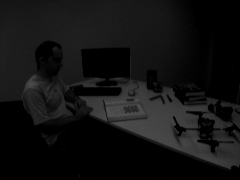}}}
    \hspace{6pt}%
    \subfloat[Image Reconstruction]{\stackinset{r}{0pt}{b}{0pt}{\frame{\includegraphics[width=45pt]{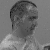}}}{\frame{\includegraphics[width=\exvis]{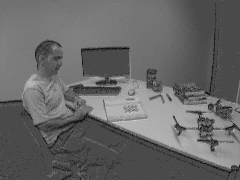}}}}
    % \vspace{-6pt}%
    \\
    \subfloat[Encrypted Events]{\frame{\includegraphics[width=\exvis]{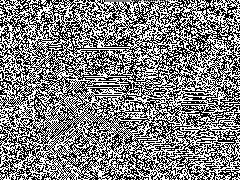}}}
    \hspace{6pt}%
    \subfloat[Event-only Reconstruction]{\frame{\includegraphics[width=\exvis]{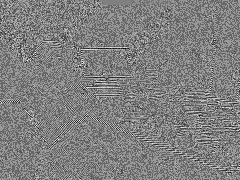}}}
    \hspace{6pt}%
    \subfloat[Failed Image Reconstruction]{\stackinset{r}{0pt}{b}{0pt}{\frame{\includegraphics[width=45pt]{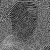}}}{\frame{\includegraphics[width=\exvis]{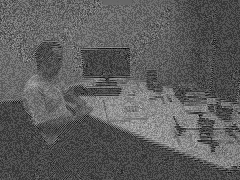}}}}
    \hspace{6pt}%
    \caption{Attacks from grid-based visualization of event streams. (a) The raw events from a neuromorphic camera can capture some dark details that are lost in (b) the underexposed image. (c) The CF~\cite{scheerlinck2018continuous} bridges (a) and (b) to reconstruct an image with high-dynamic-range states, where sensitive face features (with a zoom-in view) are recovered and exposed. (d) The encrypted events processed by our algorithm conceal all the information. (e) The failed reconstruction by the CF when only the encrypted events are used. (f) The CF uses (d) and (b) to reconstruct an image, where the face features are obscured and protected.}
    \label{fig:ex_vis}
\end{figure*}

%% file: src/ex_denoising.tex
\begin{figure}[t]
    \centering
    \includegraphics[height=\exdenoising]{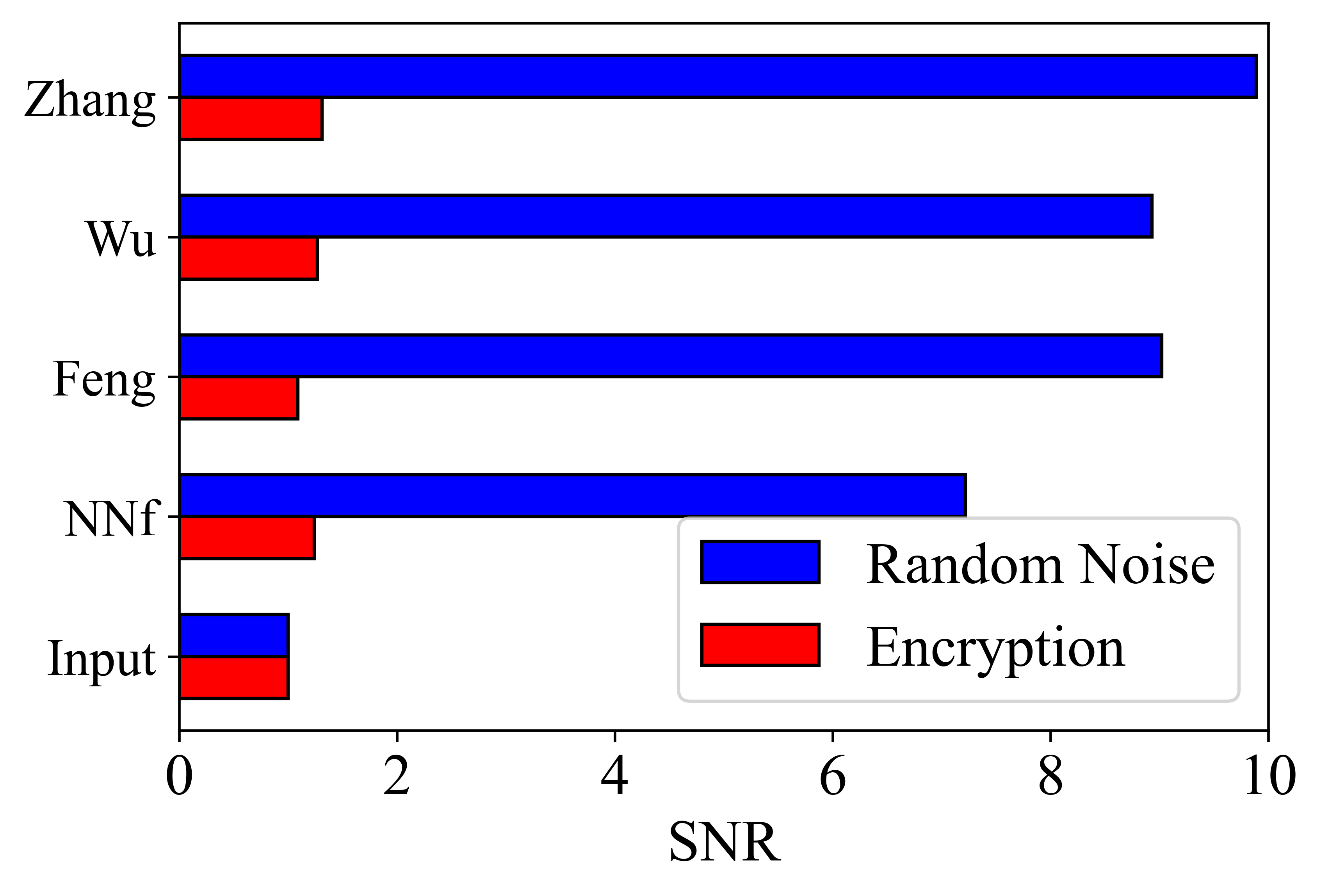}
    \vspace{-10pt}
    \caption{SNR (in ratio) comparisons on denoising two kinds of event data.}
    \label{fig:ex_stats}
\end{figure}

%% file: src/ex_recognition.tex
\begin{table}[t]
\centering
\caption{Recognition accuracy on raw events, on \underline{raw events with $50\%$ random noise}, on \textbf{the encrypted events by the counterpart} and on \textcolor{blue}{\textbf{our encrypted events}}. The symbol $\downarrow$ represents the drop we achieve in comparison with the counterpart.}
\begin{tabular}{lc|c}
\toprule
\textbf{Method} & \multicolumn{2}{c}{\textbf{Evaluation} (Top-1 Accuracy)} \\ \midrule
& N-MNIST & ASL-DVS \\
\cmidrule(lr){2-2} \cmidrule(lr){3-3}
 Rebecq~\etal~\cite{rebecq2017real} & 0.93~~\underline{0.71}~~\textbf{0.17}~~\textcolor{blue}{\textbf{0.14}}\hspace{1pt}$\scriptstyle \downarrow$ & 0.82~~\underline{0.54}~~\textbf{0.05}~~\textcolor{blue}{\textbf{0.03}}\hspace{1pt}$\scriptstyle \downarrow$ \\ 
 Gehrig~\etal~\cite{gehrig2019end} & 0.97~~\underline{0.74}~~\textbf{0.11}~~\textcolor{blue}{\textbf{0.09}}\hspace{1pt}$\scriptstyle \downarrow$ & 0.85~~\underline{0.58}~~\textbf{0.09}~~\textcolor{blue}{\textbf{0.04}}\hspace{1pt}$\scriptstyle \downarrow$ \\ 
 Li~\etal~\cite{li2021graph} & 0.95~~\underline{0.77}~~\textbf{0.15}~~\textcolor{blue}{\textbf{0.06}}\hspace{1pt}$\scriptstyle \downarrow$ & 0.88~~\underline{0.60}~~\textbf{0.11}~~\textcolor{blue}{\textbf{0.03}}\hspace{1pt}$\scriptstyle \downarrow$ \\ 
 Wang~\etal~\cite{wang2022exploiting} & 0.94~~\underline{0.68}~~\textbf{0.19}~~\textcolor{blue}{\textbf{0.12}}\hspace{1pt}$\scriptstyle \downarrow$ & 0.86~~\underline{0.55}~~\textbf{0.07}~~\textcolor{blue}{\textbf{0.06}}\hspace{1pt}$\scriptstyle \downarrow$\\ \midrule
 Random Guess & 0.1 & 0.04\\
\bottomrule
\end{tabular}
\label{tab:ex_recognition}
\end{table}

%% file: src/ex_feature.tex
\begin{figure}[t]
    \centering
    \subfloat[]{\includegraphics[height=\exdenoising]{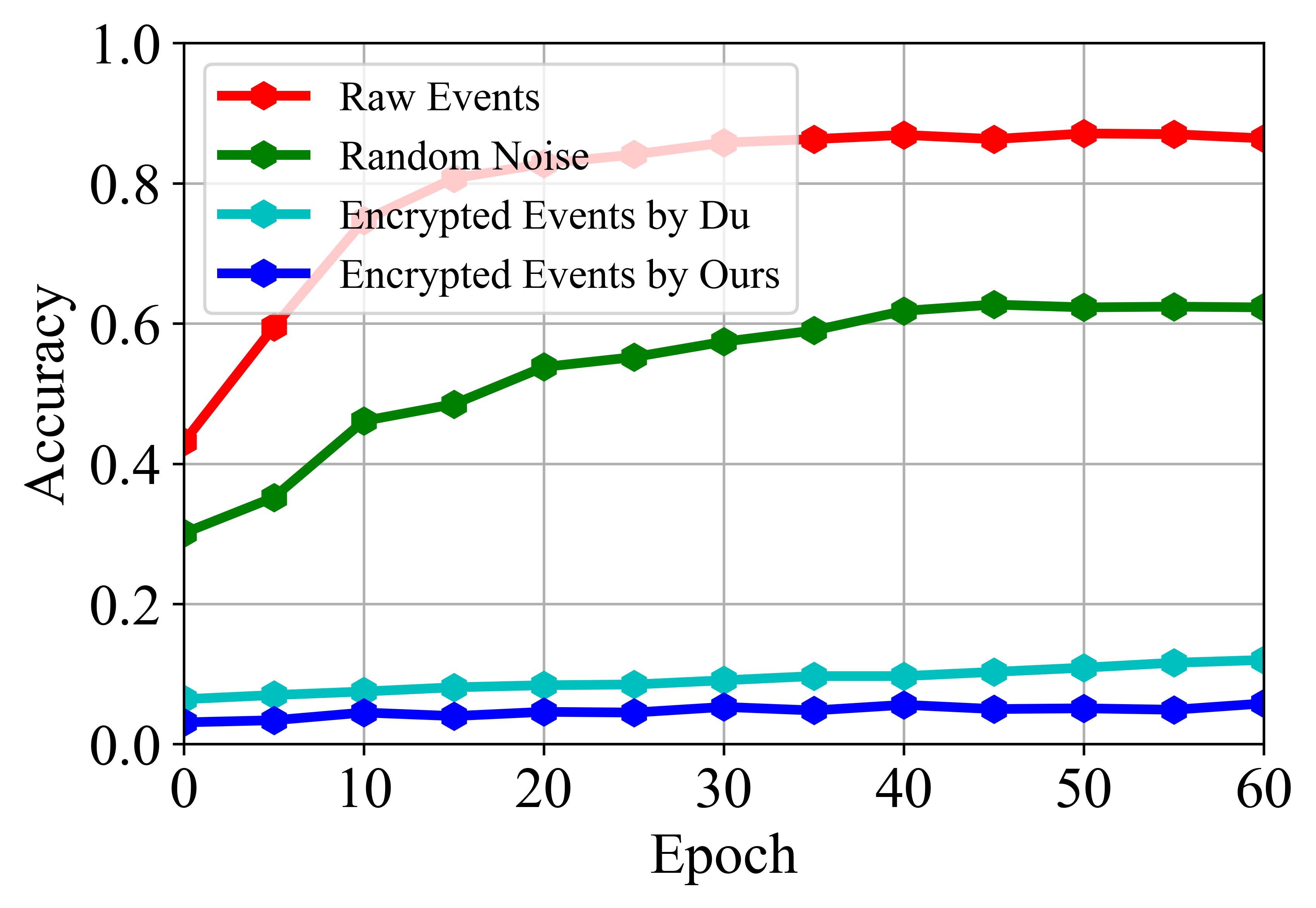}}\hspace{10pt}
    \begin{tabular}[b]{c}%
            \subfloat[]{\frame{\includegraphics[width=\exfeature]{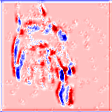}}}\\
        \subfloat[]{\frame{\includegraphics[width=\exfeature]{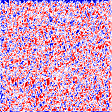}}}
    \end{tabular}
    \caption{(a) Training accuracy that changes with the increase of epochs. When a network makes inference on event frames, the feature responses of (b) raw and (c) encrypted events. The sample is from \texttt{Letter A} of ASL-DVS.}
    \label{fig:ex_feature}
\end{figure}

%% file: src/ex_detect.tex
\begin{table}[t]
\centering
\caption{Detection precision on raw events, on \underline{raw events with $50\%$ random noise}, on \textbf{the encrypted events by the counterpart} and on \textcolor{red}{\textbf{our encrypted events}}. The symbol $\downarrow$ represents the drop we achieve in comparison with the counterpart.}
\begin{tabular}{lc|c}
\toprule
\textbf{Method} & \multicolumn{2}{c}{\textbf{Evaluation} (Mean Average Precision~\cite{lin2014microsoft})} \\ \midrule
& 1 MEGAPIXEL & GEN1 \\
\cmidrule(lr){2-2} \cmidrule(lr){3-3}
 YOLOv3E~\cite{jiang2019mixed} & 34.6~~\underline{13.2}~~\textbf{3.1}~~\textcolor{red}{\textbf{1.4}}\hspace{1pt}$\scriptstyle \downarrow$ & 31.2~~\underline{18.8}~~\textbf{2.0}~~\textcolor{red}{\textbf{1.4}}\hspace{1pt}$\scriptstyle \downarrow$ \\ 
 RVT-T~\cite{gehrig2023recurrent} & 41.5~~\underline{16.4}~~\textbf{5.1}~~\textcolor{red}{\textbf{2.9}}\hspace{1pt}$\scriptstyle \downarrow$ & 44.1~~\underline{23.1}~~\textbf{3.6}~~\textcolor{red}{\textbf{3.2}}\hspace{1pt}$\scriptstyle \downarrow$ \\ 
 RVT-B~\cite{gehrig2023recurrent} & 47.4~~\underline{22.1}~~\textbf{3.3}~~\textcolor{red}{\textbf{2.5}}\hspace{1pt}$\scriptstyle \downarrow$ & 47.2~~\underline{25.7}~~\textbf{1.4}~~\textcolor{red}{\textbf{1.3}}\hspace{1pt}$\scriptstyle \downarrow$ \\ 
\bottomrule
\end{tabular}
\label{tab:ex_detect}
\end{table}

%% file: src/ex_detection.tex
\begin{figure}[t]
    \centering
    \subfloat[]{\frame{\includegraphics[width=\exdetection]{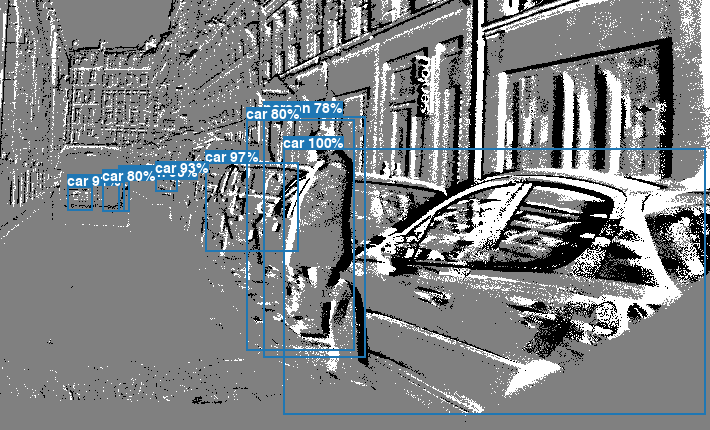}}}\hspace{2pt}
    \subfloat[]{\frame{\includegraphics[width=\exdetection]{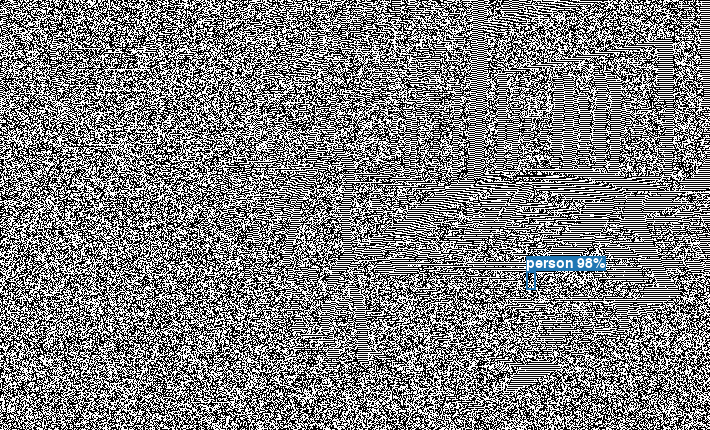}}}
    \caption{Pedestrian detection results of (a) raw events and (b) our encrypted events. The sample is from 1 MEGAPIXEL Dataset.}
    \label{fig:ex_detection}
\end{figure}

%% file: paper_NCE.bbl
\begin{thebibliography}{10}
\providecommand{\url}[1]{#1}
\csname url@samestyle\endcsname
\providecommand{\newblock}{\relax}
\providecommand{\bibinfo}[2]{#2}
\providecommand{\BIBentrySTDinterwordspacing}{\spaceskip=0pt\relax}
\providecommand{\BIBentryALTinterwordstretchfactor}{4}
\providecommand{\BIBentryALTinterwordspacing}{\spaceskip=\fontdimen2\font plus
\BIBentryALTinterwordstretchfactor\fontdimen3\font minus \fontdimen4\font\relax}
\providecommand{\BIBforeignlanguage}[2]{{%
\expandafter\ifx\csname l@#1\endcsname\relax
\typeout{** WARNING: IEEEtran.bst: No hyphenation pattern has been}%
\typeout{** loaded for the language `#1'. Using the pattern for}%
\typeout{** the default language instead.}%
\else
\language=\csname l@#1\endcsname
\fi
#2}}
\providecommand{\BIBdecl}{\relax}
\BIBdecl

\bibitem{gallego2020event}
G.~Gallego, T.~Delbruck, G.~M. Orchard, C.~Bartolozzi, B.~Taba, A.~Censi, S.~Leutenegger, A.~Davison, J.~Conradt, K.~Daniilidis, and D.~Scaramuzza, ``Event-based vision: A survey,'' \emph{IEEE Transactions on Pattern Analysis and Machine Intelligence}, vol.~44, no.~1, pp. 154--180, January 2022.

\bibitem{belbachir2012care}
A.~N. Belbachir, M.~Litzenberger, S.~Schraml, M.~Hofst{\"a}tter, D.~Bauer, P.~Sch{\"o}n, M.~Humenberger, C.~Sulzbachner, T.~Lunden, and M.~Merne, ``Care: A dynamic stereo vision sensor system for fall detection,'' in \emph{IEEE International Symposium on Circuits and Systems}, May 2012, pp. 731--734.

\bibitem{du2021event}
B.~Du, W.~Li, Z.~Wang, M.~Xu, T.~Gao, J.~Li, and H.~Wen, ``Event encryption for neuromorphic vision sensors: Framework, algorithm, and evaluation,'' \emph{Sensors}, vol.~21, no.~13, p. 4320, 2021.

\bibitem{lichtsteiner2008128}
P.~Lichtsteiner, C.~Posch, and T.~Delbruck, ``A $128 \times 128$ $120$ \si{\decibel} $15$ \si{\us} latency asynchronous temporal contrast vision sensor,'' \emph{IEEE Journal of Solid-State Circuits}, vol.~43, no.~2, pp. 566--576, February 2008.

\bibitem{taverni2018front}
G.~Taverni, D.~P. Moeys, C.~Li, C.~Cavaco, V.~Motsnyi, D.~S.~S. Bello, and T.~Delbruck, ``Front and back illuminated dynamic and active pixel vision sensors comparison,'' \emph{IEEE Transactions on Circuits and Systems II: Express Briefs}, vol.~65, no.~5, pp. 677--681, May 2018.

\bibitem{ge2023millisecond}
Z.~Ge, H.~Wei, F.~Xu, Y.~Gao, Z.~Chu, H.~K.-H. So, and E.~Y. Lam, ``Millisecond autofocusing microscopy using neuromorphic event sensing,'' \emph{Optics and Lasers in Engineering}, vol. 160, pp. 107\,247(1–--9), January 2023.

\bibitem{shiba22eccv}
S.~Shiba, Y.~Aoki, and G.~Gallego, ``Secrets of event-based optical flow,'' in \emph{European Conference on Computer Vision}, October 2022, pp. 628--645.

\bibitem{ge2022lens}
Z.~Ge, P.~Zhang, Y.~Gao, H.~K.-H. So, and E.~Y. Lam, ``Lens-free motion analysis via neuromorphic laser speckle imaging,'' \emph{Optics Express}, vol.~30, no.~2, pp. 2206--2218, January 2022.

\bibitem{zhang2022unifying}
X.~Zhang and L.~Yu, ``Unifying motion deblurring and frame interpolation with events,'' in \emph{Proceedings of the IEEE/CVF Conference on Computer Vision and Pattern Recognition}, June 2022, pp. 17\,765--17\,774.

\bibitem{becattini2022understanding}
F.~Becattini, F.~Palai, and A.~Del~Bimbo, ``Understanding human reactions looking at facial microexpressions with an event camera,'' \emph{IEEE Transactions on Industrial Informatics}, vol.~18, no.~12, pp. 9112--9121, December 2022.

\bibitem{wang2019ev}
Y.~Wang, B.~Du, Y.~Shen, K.~Wu, G.~Zhao, J.~Sun, and H.~Wen, ``{EV-Gait}: Event-based robust gait recognition using dynamic vision sensors,'' in \emph{Proceedings of the IEEE/CVF Conference on Computer Vision and Pattern Recognition}, June 2019, pp. 6358--6367.

\bibitem{guo2022low}
S.~Guo and T.~Delbruck, ``Low cost and latency event camera background activity denoising,'' \emph{IEEE Transactions on Pattern Analysis and Machine Intelligence}, vol.~45, no.~1, pp. 785--795, January 2023.

\bibitem{gehrig2019end}
D.~Gehrig, A.~Loquercio, K.~G. Derpanis, and D.~Scaramuzza, ``End-to-end learning of representations for asynchronous event-based data,'' in \emph{Proceedings of IEEE International Conference on Computer Vision}, October 2019, pp. 5633--5643.

\bibitem{perot2020learning}
E.~Perot, P.~de~Tournemire, D.~Nitti, J.~Masci, and A.~Sironi, ``Learning to detect objects with a 1 megapixel event camera,'' in \emph{Proceedings of the 34th International Conference on Neural Information Processing Systems}, vol.~33, December 2020, pp. 16\,639--16\,652.

\bibitem{behnia2008novel}
S.~Behnia, A.~Akhshani, H.~Mahmodi, and A.~Akhavan, ``A novel algorithm for image encryption based on mixture of chaotic maps,'' \emph{Chaos, Solitons \& Fractals}, vol.~35, no.~2, pp. 408--419, January 2008.

\bibitem{hou2022image}
J.~Hou and G.~Situ, ``Image encryption using spatial nonlinear optics,'' \emph{eLight}, vol.~2, no.~1, p.~3, 2022.

\bibitem{qu2020reprogrammable}
G.~Qu, W.~Yang, Q.~Song, Y.~Liu, C.-W. Qiu, J.~Han, D.-P. Tsai, and S.~Xiao, ``Reprogrammable meta-hologram for optical encryption,'' \emph{Nature Communications}, vol.~11, no.~1, p. 5484, 2020.

\bibitem{zhao2022deep}
Y.~Zhao, S.~Zheng, and X.~Yuan, ``Deep equilibrium models for video snapshot compressive imaging,'' \emph{arXiv preprint arXiv:2201.06931}, 2022.

\bibitem{cambareri2015low}
V.~Cambareri, M.~Mangia, F.~Pareschi, R.~Rovatti, and G.~Setti, ``Low-complexity multiclass encryption by compressed sensing,'' \emph{IEEE Transactions on Signal Processing}, vol.~63, no.~9, pp. 2183--2195, May 2015.

\bibitem{khodamoradi2018n}
A.~Khodamoradi and R.~Kastner, ``${O}(n)$-space spatiotemporal filter for reducing noise in neuromorphic vision sensors,'' \emph{IEEE Transactions on Emerging Topics in Computing}, vol.~9, no.~1, pp. 15--23, January 2018.

\bibitem{czech2016evaluating}
D.~Czech and G.~Orchard, ``Evaluating noise filtering for event-based asynchronous change detection image sensors,'' in \emph{IEEE International Conference on Biomedical Robotics and Biomechatronics}, June 2016, pp. 19--24.

\bibitem{feng2020event}
Y.~Feng, H.~Lv, H.~Liu, Y.~Zhang, Y.~Xiao, and C.~Han, ``Event density based denoising method for dynamic vision sensor,'' \emph{Applied Sciences}, vol.~10, no.~6, p. 2024, March 2020.

\bibitem{szudzik2006elegant}
M.~Szudzik, ``An elegant pairing function,'' in \emph{Wolfram Research (ed.) Special NKS 2006 Wolfram Science Conference}, 2006, pp. 1--12.

\bibitem{mueggler2017event}
E.~Mueggler, H.~Rebecq, G.~Gallego, T.~Delbruck, and D.~Scaramuzza, ``The event-camera dataset and simulator: Event-based data for pose estimation, visual odometry, and {SLAM},'' \emph{The International Journal of Robotics Research}, vol.~36, no.~2, pp. 142--149, February 2017.

\bibitem{orchard2015converting}
G.~Orchard, A.~Jayawant, G.~K. Cohen, and N.~Thakor, ``Converting static image datasets to spiking neuromorphic datasets using saccades,'' \emph{Frontiers in Neuroscience}, vol.~9, p. 437, October 2015.

\bibitem{bi2019graph}
Y.~Bi, A.~Chadha, A.~Abbas, E.~Bourtsoulatze, and Y.~Andreopoulos, ``Graph-based object classification for neuromorphic vision sensing,'' in \emph{Proceedings of IEEE International Conference on Computer Vision}, October 2019, pp. 491--501.

\bibitem{de2020large}
P.~De~Tournemire, D.~Nitti, E.~Perot, D.~Migliore, and A.~Sironi, ``A large scale event-based detection dataset for automotive,'' \emph{arXiv preprint arXiv:2001.08499}, 2020.

\bibitem{scheerlinck2018continuous}
C.~Scheerlinck, N.~Barnes, and R.~Mahony, ``Continuous-time intensity estimation using event cameras,'' in \emph{Asian Conference on Computer Vision}, December 2018, pp. 308--324.

\bibitem{rebecq2017real}
H.~Rebecq, T.~Horstschaefer, and D.~Scaramuzza, ``Real-time visual-inertial odometry for event cameras using keyframe-based nonlinear optimization,'' in \emph{British Machine Vision Conference}, September 2017.

\bibitem{zhang2023neuro}
P.~Zhang, Z.~Ge, L.~Song, and E.~Y. Lam, ``Neuromorphic imaging with density-based spatiotemporal denoising,'' \emph{IEEE Transactions on Computational Imaging}, vol.~9, pp. 530--541, May 2023.

\bibitem{wu2020probabilistic}
J.~Wu, C.~Ma, L.~Li, W.~Dong, and G.~Shi, ``Probabilistic undirected graph based denoising method for dynamic vision sensor,'' \emph{IEEE Transactions on Multimedia}, vol.~23, pp. 1148--1159, May 2020.

\bibitem{li2021graph}
Y.~Li, H.~Zhou, B.~Yang, Y.~Zhang, Z.~Cui, H.~Bao, and G.~Zhang, ``Graph-based asynchronous event processing for rapid object recognition,'' in \emph{Proceedings of IEEE International Conference on Computer Vision}, October 2021, pp. 934--943.

\bibitem{wang2022exploiting}
Z.~Wang, Y.~Hu, and S.-C. Liu, ``Exploiting spatial sparsity for event cameras with visual transformers,'' in \emph{International Conference on Image Processing}, October 2022, pp. 411--415.

\bibitem{lin2014microsoft}
T.-Y. Lin, M.~Maire, S.~Belongie, J.~Hays, P.~Perona, D.~Ramanan, P.~Doll{\'a}r, and C.~L. Zitnick, ``Microsoft {COCO}: Common objects in context,'' in \emph{European Conference on Computer Vision}, October 2014, pp. 740--755.

\bibitem{jiang2019mixed}
Z.~Jiang, P.~Xia, K.~Huang, W.~Stechele, G.~Chen, Z.~Bing, and A.~Knoll, ``Mixed frame-/event-driven fast pedestrian detection,'' in \emph{International Conference on Robotics and Automation}, May 2019, pp. 8332--8338.

\bibitem{gehrig2023recurrent}
M.~Gehrig and D.~Scaramuzza, ``Recurrent vision transformers for object detection with event cameras,'' in \emph{Proceedings of the IEEE/CVF Conference on Computer Vision and Pattern Recognition}, June 2023, pp. 13\,884--13\,893.

\end{thebibliography}
